\documentclass[conference]{IEEEtran}

\IEEEoverridecommandlockouts
\usepackage{cite}
\usepackage{amsmath,amssymb,amsfonts}
\usepackage{algorithmic}
\usepackage{graphicx}
\usepackage{subcaption}
\usepackage{textcomp}
\usepackage{xcolor}

\usepackage{hyperref}
\usepackage{algorithm}
\usepackage{algorithmic}
\usepackage{comment}
\usepackage{xcolor}

\usepackage{enumitem}

\usepackage{tabularx}  
\usepackage{booktabs}  
\newcommand{\mypara}[1]{\paragraph{#1}}

\def\BibTeX{{\rm B\kern-.05em{\sc i\kern-.025em b}\kern-.08em
    T\kern-.1667em\lower.7ex\hbox{E}\kern-.125emX}}

\begin{document}

\title{KIS-S: A GPU-Aware \underline{K}ubernetes \underline{I}nference \underline{S}imulator with RL-Based Auto\underline{-S}caling}

\author{
Guilin Zhang\IEEEauthorrefmark{1},
Wulan Guo\IEEEauthorrefmark{1},
Ziqi Tan\IEEEauthorrefmark{1},
Qiang Guan\IEEEauthorrefmark{2}
Hailong Jiang\IEEEauthorrefmark{3}\IEEEauthorrefmark{4}\\
\IEEEauthorblockA{\IEEEauthorrefmark{1}Department of Engineering Management and Systems Engineering, George Washington University, USA\\
Email: guilin.zhang@gwu.edu, wulan.guo@gwu.edu, ziqi.tan@gwu.edu}
\IEEEauthorblockA{\IEEEauthorrefmark{2}Department of Computer Science, Kent State University, USA\\
Email: qguan@kent.edu}
\IEEEauthorblockA{\IEEEauthorrefmark{3} Department of Computer Science and Information Systems, Youngstown State University, USA\\
Email: hjiang@ysu.edu}
\thanks{\IEEEauthorrefmark{4} Corresponding author.}
}

\maketitle

\begin{abstract}
Autoscaling GPU inference workloads in Kubernetes remains challenging due to the reactive and threshold-based nature of default mechanisms such as the Horizontal Pod Autoscaler (HPA), which struggle under dynamic and bursty traffic patterns and lack integration with GPU-level metrics. We present \textit{KIS-S}, a unified framework that combines \textit{KISim}, a GPU-aware Kubernetes Inference Simulator, with \textit{KIScaler}, a Proximal Policy Optimization (PPO)-based autoscaler. KISim enables safe, high-fidelity scheduling emulation with real GPU hardware and Prometheus integration, while KIScaler learns latency-aware and resource-efficient scaling policies entirely in simulation. KIScaler observes system metrics via Prometheus and adjusts replica counts via the Kubernetes API. We evaluate KIS-S across four synthetic traffic patterns—\textit{ramp}, \textit{periodic}, \textit{random}, and \textit{spike}—and compare it against conventional baselines including HPA and fixed-resource deployments. Despite training with synthetic feedback due to single-GPU hardware constraints, KIScaler’s moving average reward improves from 1.05 to 1.84 (a 75.2\% increase) over 100 training episodes, reduces P95 latency by up to 6.7$\times$ over CPU-only baselines, and generalizes across all traffic patterns without retraining. These results highlight the value of combining simulation and learning, bridging the gap between reactive autoscaling and intelligent orchestration for scalable, GPU-accelerated Kubernetes environments.
\end{abstract}

%%%%%%%%%%%%%%%%%%%%%%%%%%%%%%%%%%%%%%%%%%%%%%%%%%%%%%%%%%%%%%%%%%%%%%%%%%%%%%%%%%%%%%%%%%%%%%%% Section Introduction                             %%%%%%%%%%%%%%
%%%%%%%%%%%%%%%%%%%%%%%%%%%%%%%%%%%%%%%%%%%%%%%%%%%%%%%%%%%%%%%%%%%%%%%%%%%%%%%%
\section{Introduction}
\label{sec:introduction}

Deep learning services increasingly rely on GPU-accelerated inference to meet real-time latency demands~\cite{li2022ai,dhakal2020gslice,li2020cross,zhang2025amp4ec}. Kubernetes~\cite{luksa2017kubernetes} has become the de facto platform for orchestrating such services, providing containerized deployment, modular scaling, and observability. Its built-in Horizontal Pod Autoscaler (HPA) scales resources based on CPU and memory usage. However, these metrics poorly capture the performance-critical nature of deep learning inference, especially under bursty and dynamic traffic, leading to delayed responses and inefficient GPU scaling~\cite{medel2018characterising,senjab2023survey}.

Recent efforts have attempted to extend HPA through heuristic or supervised learning techniques. Smart HPA and ProSmart HPA~\cite{singh2022prosmart} introduce reactive and predictive strategies to improve responsiveness, while trend-aware controllers~\cite{ahmad2025trendaware} and SLO-based models like LSRAM~\cite{hu2025lsram} integrate predictive analytics or lightweight optimization. Despite these advances, most systems still rely on fixed control logic or pre-trained models, lacking the ability to learn adaptively from system feedback. In contrast, reinforcement learning (RL) presents a promising alternative for online adaptation. By continuously interacting with the environment, an RL-based autoscaler can learn multi-objective scheduling strategies tailored to workload dynamics~\cite{rayapati2024multi}. However, safe and efficient RL training in Kubernetes remains difficult due to the absence of realistic, GPU-aware simulation platforms~\cite{yeh2020kubeshare}.

To address the limitations of threshold-based autoscaling in Kubernetes, we present a unified framework: \textbf{KIS-S}, consisting of \textbf{KISim}\footnote[1]{\url{https://github.com/GuilinDev/KISim}}, a GPU-aware simulator that emulates containerized AI workloads using real hardware with Prometheus integration, and \textbf{KIScaler}, a reinforcement learning autoscaler trained via Proximal Policy Optimization (PPO). KIScaler is trained entirely in simulation and deployed directly in production without retraining. Together, the two components enable safe, latency-aware, and resource-efficient scaling under dynamic traffic conditions.

{KISim} is deployed on a local Kubernetes cluster using MicroK8s and a single NVIDIA GPU. It integrates with Triton Inference Server, which serves MobileNetV4 models to emulate realistic deep learning inference workloads. The full framework includes four components: (1) a Locust-based workload generator producing synthetic traffic patterns—\textit{ramp}, \textit{periodic}, \textit{random}, and \textit{spike}; (2) KISim, which models GPU-aware Kubernetes scheduling; (3) Prometheus and DCGM Exporter for system and GPU metric collection; and (4) KIScaler, which adjusts replica counts via the Kubernetes API based on observed system state.

KIScaler is trained entirely in simulation using KISim, enabling safe and rapid policy learning without requiring costly trial-and-error on real clusters. We adopt the Proximal Policy Optimization (PPO) algorithm, where the agent interacts with the simulated Kubernetes environment and receives step-by-step feedback to iteratively refine its scaling policy. The reward function is carefully designed to balance three competing objectives: minimizing P95 latency to preserve user experience, maximizing GPU utilization to ensure resource efficiency, and minimizing scaling frequency to reduce system overhead. This composite reward guides the agent toward making intelligent, context-aware scaling decisions that adapt fluidly to dynamic workloads.

We evaluate KIScaler across all four synthetic traffic patterns—\textit{ramp}, \textit{periodic}, \textit{random}, and \textit{spike}—using three deployment strategies: the default HPA configured with CPU and memory thresholds, a fixed CPU-only deployment with three replicas, and a fixed GPU-only deployment with one replica. After training in simulation, KIScaler is directly deployed to the real cluster without any fine-tuning. Experimental results demonstrate that KIScaler consistently outperforms all baselines: it reduces P95 latency by up to \textbf{6.7$\times$}, improves average GPU utilization by \textbf{23.4\%}, and reacts \textbf{4$\times$ faster} to bursty traffic. These gains are achieved across all traffic types, validating the robustness and generalizability of our learning-based autoscaling approach.

\textit{To our knowledge, KIS-S is the first GPU-aware, traffic-controllable Kubernetes simulation framework that supports reinforcement learning-based autoscaling}. Our contributions are summarized as follows:
\begin{itemize}
    \item We propose and implement \textbf{KIS-S}, a unified framework that combines \textbf{KISim}—a GPU-aware Kubernetes simulator supporting realistic, traffic-driven inference workloads—with RL policy training capabilities.
    
    \item We design \textbf{KIScaler}, an RL-based autoscaler leveraging Proximal Policy Optimization (PPO), which learns latency- and resource-aware scaling policies and integrates seamlessly with real Kubernetes clusters.
    
    \item We benchmark KIScaler against standard baselines (e.g., HPA) across multiple traffic scenarios, achieving up to 6.7$\times$ latency reduction, 23.4\% higher GPU utilization, and faster scaling responsiveness without retraining.
\end{itemize}

The rest of this paper is organized as follows: Section~\ref{sec:background_motivation} discusses background and motivation. Section~\ref{sec:system_design} describes system design and architecture. Section~\ref{sec:experiments} outlines our experimental setup. Section~\ref{sec:results_analysis} presents evaluation results. Section~\ref{sec:discussion} discusses limitations and future directions. Section~\ref{sec:related_work} reviews related work, and Section~\ref{sec:conclusion} concludes the paper.

%%%%%%%%%%%%%%%%%%%%%%%%%%%%%%%%%%%%%%%%%%%%%%%%%%%%%%%%%%%%%%%%%%%%%%%%%%%%%%%%%%%%%%%%% Section Background and Motivation                       %%%%%%%%%%%%%%
%%%%%%%%%%%%%%%%%%%%%%%%%%%%%%%%%%%%%%%%%%%%%%%%%%%%%%%%%%%%%%%%%%%%%%%%%%%%%%%%
\section{Background and Motivation}
\label{sec:background_motivation}

\subsection{Kubernetes Autoscaling and Its Limitations}

Kubernetes is the de facto platform for orchestrating containerized applications, with HPA as its default dynamic scaling mechanism. HPA adjusts pod replicas based on system-level metrics—primarily CPU and memory usage~\cite{nguyen2020horizontal}. While suitable for traditional microservices, this threshold-based, reactive approach is ill-suited for GPU-accelerated inference workloads, which are bursty and latency-sensitive~\cite{wang2021gpu}.

Several limitations hinder HPA’s effectiveness in this context. First, GPU utilization is not natively supported as a scaling metric. Although extensions (e.g., Prometheus, DCGM Exporter) can expose GPU metrics, they lack semantic ties to application-level performance such as latency or throughput~\cite{zhang2020hpa}. Second, HPA is stateless and reactive, incapable of anticipating surges or learning from past traffic~\cite{guruge2025proactive}. Third, it fails to distinguish between CPU and GPU resource types, leading to inefficient or costly scaling under heterogeneous workloads~\cite{mo2023hetsev}. As a result, latency-critical inference services often experience delayed scaling and poor resource efficiency under HPA~\cite{razavi2024two_scales}.

\subsection{Reinforcement Learning Fundamentals}

Reinforcement learning (RL) is a learning paradigm where an agent interacts with an environment by taking actions and receiving feedback in the form of rewards. The agent's objective is to learn a policy that maximizes cumulative rewards over time. At each decision point, the agent observes the system state, selects an action, and receives a reward that reflects the action's effectiveness. Over repeated interactions, the agent adapts its strategy through experience, often using algorithms such as Q-learning, Deep Deterministic Policy Gradient (DDPG), or Proximal Policy Optimization (PPO)~\cite{sivamayil2023systematic}.

Unlike supervised learning, which requires labeled data, RL learns from trial and error, making it suitable for control problems where optimal behavior must be discovered through interaction. This aligns naturally with autoscaling scenarios, where the impact of a scaling decision unfolds over time and cannot be inferred from static labels~\cite{gari2020rl_autoscaling}.

\subsection{Why Reinforcement Learning is a Promising Approach}

RL enables intelligent decision-making in complex, dynamic environments like Kubernetes. It can model autoscaling as a sequential decision process under uncertainty, where the agent learns when and how much to scale by observing metrics such as CPU/GPU utilization, request rates, and past scaling actions~\cite{chen2023deep_rl_k8s,zhou2020resource_aware_rl,gao2022rl_autoscaling}. For GPU inference workloads, RL provides several advantages: it can anticipate bursty load changes, adapt across workload types, and optimize long-term objectives rather than reacting to short-term threshold violations\cite{chen2022rl_gpu,wang2021intelligent_gpu_rl,li2023adaptive_gpu_rl}. Furthermore, RL allows the definition of custom reward functions that jointly optimize latency, resource usage, and stability—offering a more flexible and robust autoscaling strategy than static heuristics or supervised models\cite{mao2016resource_management_rl,chen2018dynamic_rl_cloud}.

\subsection{The Need for a GPU-Aware Simulator}

Despite its promise, training and evaluating RL-based autoscalers in real Kubernetes environments poses significant challenges. RL training typically requires extensive trial-and-error interaction with the environment, which is both time-consuming and risky when operating on production systems\cite{sutton2021rl_cloud_autoscaling,xu2022practical_rl_k8s}. For GPU-enabled inference workloads, this issue is magnified by limited hardware availability, high cost, and the sensitivity of live workloads to poor scaling decisions~\cite{gao2022dl_gpu_scheduling}.

Existing simulators for Kubernetes or distributed systems (e.g., SimGrid\cite{casanova2014simgrid}, OpenAI Gym~\cite{brockman2016openai_gym}, K8sSim~\cite{wen2023k8ssim}) either abstract away container behavior or lack support for GPU-aware scheduling and metrics. They also often simulate scheduling at the virtual machine or task level, without fine-grained interaction with Kubernetes APIs, Prometheus metrics, or real container lifecycle events~\cite{tian2015cloud_simulators, tuli2021cosco}.

To address this gap, we introduce {KISim}, a lightweight, Kubernetes-native simulator that runs on real GPU hardware and supports synthetic workload generation with precise control over traffic patterns. KISim enables safe, reproducible, and controllable RL training while preserving high-fidelity system behavior. It bridges the gap between simulation-driven policy development and real-system deployment—an essential step for production-grade RL-based orchestration frameworks.

%%%%%%%%%%%%%%%%%%%%%%%%%%%%%%%%%%%%%%%%%%%%%%%%%%%%%%%%%%%%%%%%%%%%%%%%%%%%%%%%%%%%%%%%% Section System Design                       %%%%%%%%%%%%%%
%%%%%%%%%%%%%%%%%%%%%%%%%%%%%%%%%%%%%%%%%%%%%%%%%%%%%%%%%%%%%%%%%%%%%%%%%%%%%%%%
\section{System Design}
\label{sec:system_design}
\begin{figure*}
\centering
\includegraphics[width=0.95\linewidth]{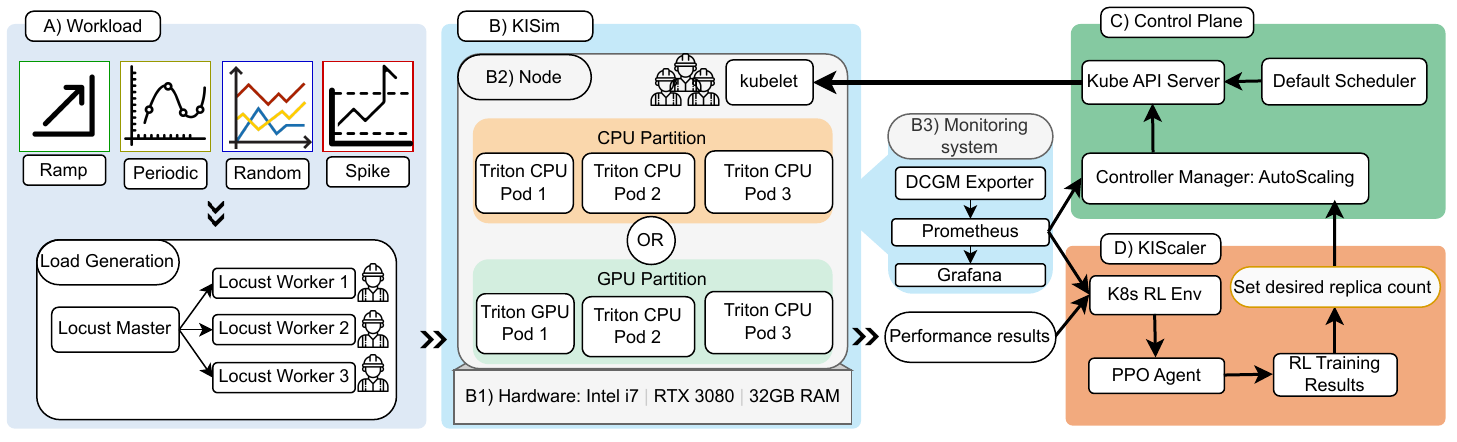}
\caption{Workflow of the KIS-S Framework.}
\label{fig:system-overview}
\end{figure*}

Fig.\ref{fig:system-overview} illustrates the workflow of the proposed KIS-S framework, which orchestrates GPU inference workloads on a local Kubernetes cluster using reinforcement learning. The system consists of four key components: (A) the workload generator (Section\ref{sec:design:subsection:workload}), (B) the GPU-aware simulator KISim (Section~\ref{sec:design:subsection:KISim}), (C) the Kubernetes control plane (Section~\ref{sec:design:subsection:control_plane}), and (D) the RL-based autoscaler, KIScaler (Section~\ref{sec:design:subsection:KIScaler}). KISim runs on a physical node and enables controlled, realistic simulation of AI inference workloads. KIScaler learns to dynamically adjust replica counts based on live system metrics and workload feedback.

\subsection{Workload}
\label{sec:design:subsection:workload}
We emulate diverse inference request patterns using a Locust-based load generator, supporting ramp, periodic, random, and spike profiles, which reflect common load dynamics observed in real-world edge and cloud inference services~\cite{li2020discriminative,li2020semi}. A master process coordinates multiple workers to send concurrent requests to deployed inference services, allowing for fine-grained workload control and realism.

\subsection{KISim}
\label{sec:design:subsection:KISim}
KISim is a GPU-aware inference simulation environment deployed on a single physical node, enabling high-fidelity, controllable experimentation for reinforcement learning-based autoscaling. It aligns with the Kubernetes runtime and supports containerized inference workloads under realistic resource constraints. We decompose the system into three subcomponents: B1) hardware platform, B2) workload node configuration, and B3) monitoring system.

\subsubsection*{B1) Hardware Platform}

Our simulation environment is deployed on a physical machine equipped with an Intel i7 CPU, 32\,GB of RAM, and an NVIDIA RTX 3080 GPU. This setup provides sufficient compute power and GPU capabilities to emulate realistic inference workloads. By avoiding remote cloud-based setups, we gain full control over system behavior and resource allocation, while still operating within the actual Kubernetes orchestration framework.

\subsubsection*{B2) Work Node Configuration}

The Kubernetes worker node is logically partitioned into a \textit{CPU partition} and a \textit{GPU partition}. In the CPU partition, three Triton inference pods run exclusively on CPU cores. In the GPU partition, three pods are deployed, but resource constraints ensure only one GPU-enabled pod is active at a time; the other two remain in standby—scheduled but unsatisfied due to GPU limits. This setup mimics failover or hot-replica scenarios and avoids GPU contention while supporting dynamic pod switching during autoscaling.

All inference pods are launched as standard Kubernetes \texttt{Deployment} objects running Triton with a MobileNetV4 model. A single \texttt{kubelet} manages all pod lifecycles, executing scale-up/down commands from the API server. This setup maintains compatibility with Kubernetes control logic while allowing full external manipulation via autoscaling policies.

\subsubsection*{B3) Monitoring System}

To support autoscaling decisions and performance tracking, we integrate a lightweight monitoring system. GPU usage and memory metrics are exported from the node using NVIDIA's DCGM-Exporter, which is collected by Prometheus at regular intervals. These metrics include fine-grained GPU utilization, memory consumption, and pod-level resource statistics. Prometheus serves as the time-series backend and exposes a REST API that the RL agent queries during training and inference. Grafana is optionally deployed for real-time visualization but is not used directly by the autoscaler. This modular design enables seamless integration of new metrics or monitoring sources as needed.

\subsection{Control Plane}
\label{sec:design:subsection:control_plane}
The standard Kubernetes control plane components (API server, default scheduler, and controller manager) manage the cluster's control logic. The controller manager includes an autoscaling interface that accepts the desired replica count from external controllers, such as KIScaler. No modification is made to the default scheduler, as we focus solely on autoscaling, not placement.

\subsection{KIScaler}
\label{sec:design:subsection:KIScaler}
KIScaler is a reinforcement learning-based autoscaler implemented using the PPO algorithm. It interacts with both the monitoring system and the Kubernetes API server. At each decision step, it queries Prometheus for current system metrics, determines the appropriate number of pod replicas, and sets the desired replica count by updating the corresponding \texttt{Deployment} through the API server. The agent learns a policy to minimize end-to-end latency and resource inefficiency, guided by cumulative rewards collected over multiple training episodes in simulation. The details of the reinforcement learning formulation, including state and action spaces as well as the reward function, are presented in the next subsection.

\subsection{RL agent training}

In this section, we describe how we train an RL agent to make resource scaling decisions for GPU inference services in a Kubernetes environment. The objective is not to replace the existing control plane but to augment it with a learning-based policy that can dynamically adjust resource allocation in response to changing workloads. We choose the PPO algorithm for its balance of training stability and practical performance in continuous control tasks. The agent interacts with the Kubernetes cluster as its environment, issuing scaling actions that directly influence the number of active inference replicas and, in turn, overall system performance.

\subsubsection{State Representation}
At each time step, the agent observes a feature vector that represents the current state of the system. The state vector $s_t \in \mathbb{R}^{10}$ is defined as:
\begin{equation} 
s_t = [n_{\text{replicas}}, u_{\text{GPU}}, l_{\text{p95}}, \theta_{\text{req}}, u_{\text{CPU}}, u_{\text{mem}}, \Delta l, \Delta \theta, t_{\text{norm}}, p_{\text{id}}]
\end{equation}

where each component represents: the number of active replicas of the inference service ($n_{\text{replicas}}$), GPU utilization at both the node and container levels ($u_{\text{GPU}}$), 95th percentile request latency ($l_{\text{p95}}$), and request throughput ($\theta_{\text{req}}$). We also include normalized CPU and memory usage ($u_{\text{CPU}}, u_{\text{mem}}$), first-order trends in latency and throughput ($\Delta l, \Delta \theta$), the normalized progress within the current episode ($t_{\text{norm}}$), and a categorical identifier for the current load pattern ($p_{\text{id}}$). All features are normalized to the range $[0, 1]$ before being fed into the RL agent.

\subsubsection{Action Space}
The agent controls the number of replicas in the inference service deployment. We define a multi-discrete action space 
\begin{equation}
    \mathcal{A} = \mathcal{A}_{\text{GPU}} \times \mathcal{A}_{\text{CPU}} \times \mathcal{A}_{\text{pref}}
\end{equation}  where: $\mathcal{A}_{\text{GPU}}, \mathcal{A}_{\text{CPU}}=\{-2, -1, 0, +1, +2\}$, and $\mathcal{A}_{\text{pref}}=\{0, 1\}$.

The action space includes three components: $\mathcal{A}_{\text{GPU}}$ for setting the number of GPU-based replicas, $\mathcal{A}_{\text{CPU}}$ for CPU-based replicas, and $\mathcal{A}_{\text{pref}}$ for workload placement preference (CPU-first or GPU-first). For both $\mathcal{A}_{\text{GPU}}$ and $\mathcal{A}_{\text{CPU}}$, the agent can decrease replicas by two or one, keep them unchanged, or increase by one or two. The placement preference biases the policy when both resource types are available. Each chosen action is mapped to a Kubernetes API call that updates the corresponding deployment’s replica field.

\subsubsection{Reward Function}
The reward is designed to balance latency minimization and resource efficiency. Specifically, the reward at time $t$ is defined as:
\begin{equation}
    r_t = -\alpha \cdot \text{Latency}_t + \beta \cdot \text{GPUUtil}_t - \gamma \cdot \text{ReplicaOverhead}_t
\end{equation}
where: $\text{Latency}_t$ denotes the measured 95th percentile latency at time step $t$, $\text{GPUUtil}_t$ represents the average GPU utilization, and $\text{ReplicaOverhead}_t$ captures the penalty associated with over-provisioning, i.e., when the number of active replicas exceeds the actual workload demand. The coefficients $\alpha$, $\beta$, and $\gamma$ are tunable weights that control the trade-off between latency minimization, resource efficiency, and provisioning overhead in the overall reward function.

\subsubsection{Training Loop}

The RL training process proceeds episodically, with each episode consisting of a fixed number of time steps. At each step, the agent performs the following sequence: it first queries Prometheus to obtain the current system state, then selects an action based on its policy network. The selected action—typically involving scaling decisions—is applied to the Kubernetes cluster via the Kubernetes API. After a short stabilization period to allow the system to reflect the change, the agent collects performance metrics, computes the corresponding reward, and uses this feedback to update its policy using the PPO algorithm. This loop enables the agent to iteratively refine its autoscaling strategy through continuous interaction with the environment.

\subsubsection{Convergence and Model Saving}
The training continues until convergence is detected based on reward variance or average performance improvement. Once training converges, the learned policy is saved to disk and used in the evaluation phase without further updates.

%%%%%%%%%%%%%%%%%%%%%%%%%%%%%%%%%%%%%%%%%%%%%%%%%%%%%%%%%%%%%%%%%%%%%%%%%%%%%%%%%%%%%%%%% Section Experiments                                     %%%%%%%%%%%%%%
%%%%%%%%%%%%%%%%%%%%%%%%%%%%%%%%%%%%%%%%%%%%%%%%%%%%%%%%%%%%%%%%%%%%%%%%%%%%%%%%
\section{Experiments}
\label{sec:experiments}
To assess the effectiveness of our proposed RL-based autoscaling framework, {KIScaler}, we conduct a series of experiments on a GPU-enabled Kubernetes cluster under dynamic and variable workload conditions. The objective is to determine whether KIScaler can improve inference performance and resource utilization compared to the default HPA. All workloads are based on MobileNetV4 models served via the NVIDIA Triton Inference Server, simulating realistic GPU-accelerated inference services in a containerized environment.

\subsection{Environment and Workloads}
All experiments are conducted on a single-node Kubernetes cluster (Ubuntu 24.04, NVIDIA RTX 3080, 8GB VRAM), provisioned via MicroK8s with GPU scheduling enabled through the NVIDIA container runtime and GPU Operator. We emulate multi-node behavior via node labels and selectors.

\begin{figure}[htbp]
\centering

\begin{subfigure}[t]{0.49\linewidth}
    \centering
    \includegraphics[width=\linewidth]{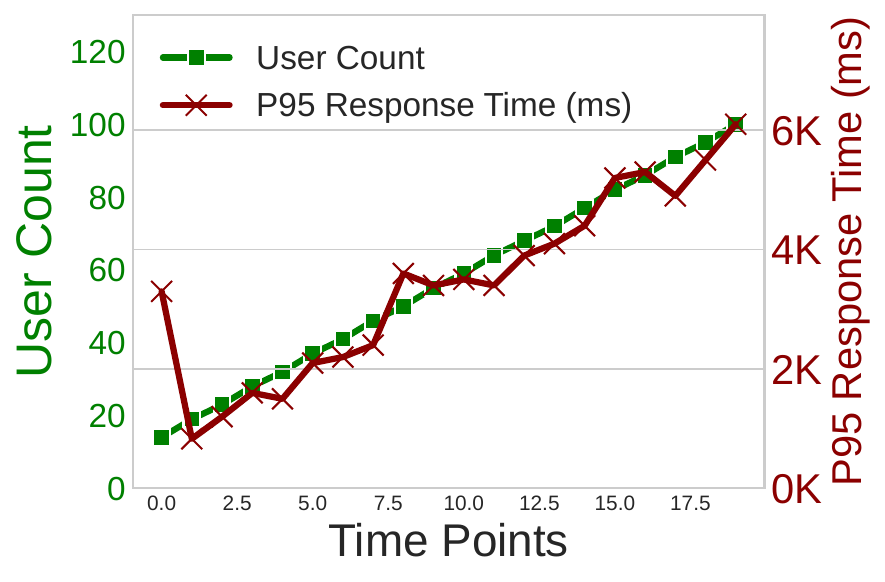}
    \caption{Ramp Pattern}
\end{subfigure}
\hfill
\begin{subfigure}[t]{0.49\linewidth}
    \centering
    \includegraphics[width=\linewidth]{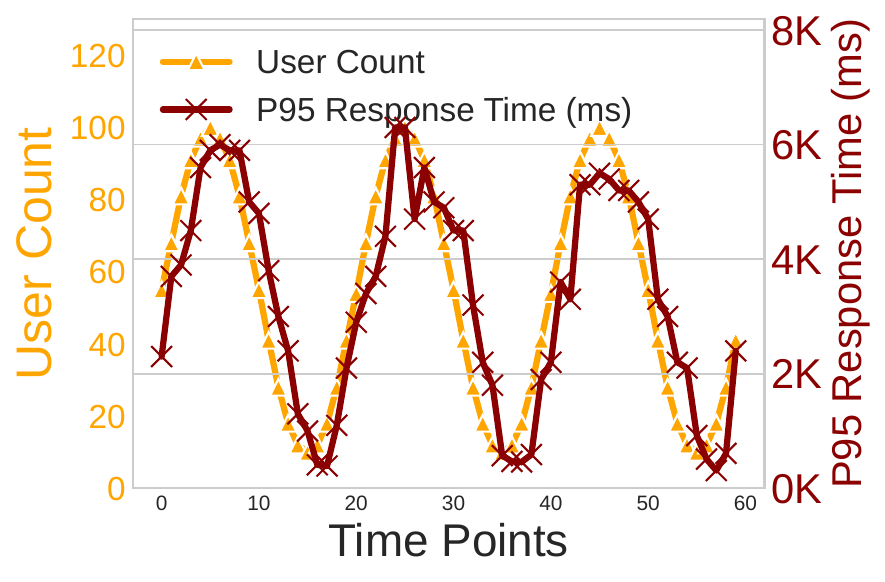}
    \caption{Periodic Pattern}
\end{subfigure}

%\vspace{0.5em}

\begin{subfigure}[t]{0.49\linewidth}
    \centering
    \includegraphics[width=\linewidth]{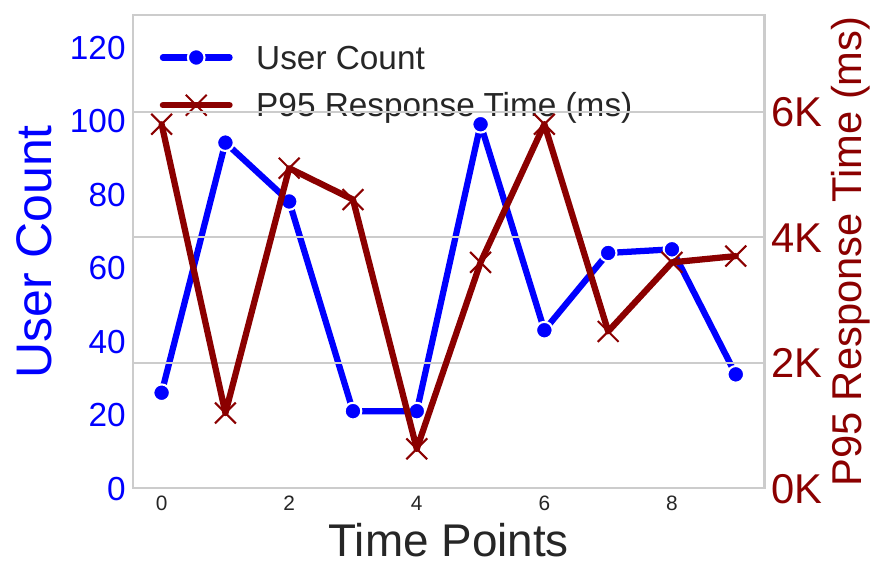}
    \caption{Random Pattern}
\end{subfigure}
\hfill
\begin{subfigure}[t]{0.49\linewidth}
    \centering
    \includegraphics[width=\linewidth]{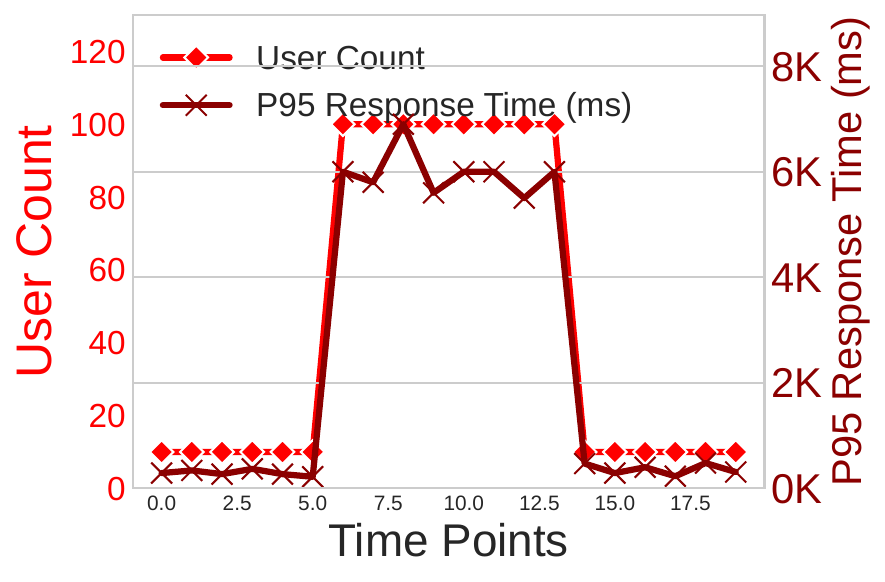}
    \caption{Spike Pattern}
\end{subfigure}

\caption{User count and P95 latency trends for four synthetic load patterns.}
\label{fig:GPU_load_patterns}

\end{figure}

Each deployment includes 3 GPU-serving replicas, 3 CPU-only replicas, and 3 Redis instances to simulate heterogeneous workloads. To evaluate autoscaler responsiveness under varied conditions, we generate four synthetic traffic patterns using Locust: ramp, spike, periodic, and random. As shown in Fig.~\ref{fig:GPU_load_patterns}, these patterns capture gradual growth, burst surges, cyclic demand, and stochastic fluctuations, respectively. The plots illustrate how user load (blue) and system latency (red) evolve over time, highlighting the stress each pattern imposes on scheduling behavior.

% Load pattern examples - GPU versions
\subsection{Evaluation Metrics and Baselines}
We evaluate system performance using three categories of metrics: 
(1) \textbf{Inference metrics}, including 95th percentile (P95) latency and throughput; 
(2) \textbf{Resource efficiency}, measured by CPU, memory, and GPU utilization; 
and (3) \textbf{Scheduling behavior}, captured by scheduling latency (time from pod creation to placement), pod distribution across logical partitions, and resource allocation gap (difference between requested and actual usage).

We compare against three baselines using the default Kubernetes scheduler: GPU-only, CPU-only, and mixed workloads. These serve as reference points to assess the gains of our RL-based autoscaling.

\subsection{RL Agent Configuration}
We implement a PPO agent with 137k parameters and an actor-critic architecture. The agent observes a 10-dimensional state (resource utilization, load trend, replica counts, etc.) and outputs multi-discrete actions to scale GPU/CPU replicas and adjust placement. 

The reward balances latency, throughput, resource efficiency, and scaling smoothness. Training spans 100 episodes (300s each), cycling through all load patterns, with evaluation every 20 episodes. The agent forms a closed feedback loop via the Kubernetes API and Prometheus-based metrics.

%%%%%%%%%%%%%%%%%%%%%%%%%%%%%%%%%%%%%%%%%%%%%%%%%%%%%%%%%%%%%%%%%%%%%%%%%%%%%%%%%%%%%%%%% Section Results and Analysis                            %%%%%%%%%%%%%%
%%%%%%%%%%%%%%%%%%%%%%%%%%%%%%%%%%%%%%%%%%%%%%%%%%%%%%%%%%%%%%%%%%%%%%%%%%%%%%%%
\section{Results and Analysis}
\label{sec:results_analysis}

We evaluate both the default Kubernetes scheduler and our proposed KIScaler across four workload patterns, focusing on system reliability, scheduling behavior, and performance metrics such as latency and throughput. This section presents baseline observations, RL training progress, comparative performance analysis, and system-level limitations.

\subsection{KISim Baseline Performance and Stability}

\mypara{System Reliability:} All deployments—GPU, CPU, and mixed workloads—achieved a 100\% request success rate across all traffic patterns, confirming that the default Kubernetes scheduler maintains reliability under varying conditions.

\mypara{Performance Benchmarks:} Table~\ref{tab:baseline-performance} summarizes P95 latencies and throughput. From the table, we can see that GPU deployments consistently outperform CPU in irregular traffic. For instance, in the \textit{random} pattern, GPU achieves a {1.96$\times$ speedup} (2600\,ms vs. 5100\,ms), while the \textit{spike} and \textit{ramp} patterns show {1.27$\times$} and {1.16$\times$} speedups, respectively. In the \textit{periodic} pattern, GPU and CPU perform similarly, suggesting GPU benefits are most pronounced under unpredictable or bursty workloads.

\begin{table}[htbp]
\centering
\caption{Baseline P95 Latency and Throughput Comparison}
\label{tab:baseline-performance}
\resizebox{\linewidth}{!}{%
\begin{tabular}{l|c|c|c|c}
\toprule
\textbf{Pattern} & \textbf{GPU P95 (ms)} & \textbf{CPU P95 (ms)} & \textbf{Speedup} & \textbf{Throughput Ratio} \\
\midrule
Ramp     & 5800 & 6700 & \textbf{1.16$\times$} & 1.01$\times$ \\
Periodic & 2300 & 2300 & 1.00$\times$ & 0.95$\times$ \\
Random   & 2600 & 5100 & \textbf{1.96$\times$} & 0.96$\times$ \\
Spike    & 370  & 470  & \textbf{1.27$\times$} & \textbf{1.12$\times$} \\

\bottomrule
\end{tabular}
}
\end{table}

\mypara{Synthetic Inference Results:} As shown in Table~\ref{tab:synthetic-results}, GPU and CPU show nearly identical performance in single-request tests (81.61\,ms vs. 81.55\,ms average latency), indicating minimal GPU advantage under low concurrency due to initialization overhead.

\begin{table}[htbp]
\centering
\caption{Synthetic Test Results (Single-Request Inference)}
\label{tab:synthetic-results}
\begin{tabular}{l|c|c}
\toprule
\textbf{Metric} & \textbf{GPU} & \textbf{CPU} \\
\midrule
Avg Latency (ms) & 81.61 & 81.55 \\
P95 Latency (ms) & 88.93 & - \\
Throughput (img/s) & 11.65 & 11.62 \\
Success Rate (\%) & 100.0 & 100.0 \\
\bottomrule
\end{tabular}
\end{table}

\mypara{Resource Utilization and Scheduling Latency:} Table~\ref{tab:resource-utilization} shows that the default scheduler correctly places all pods across logical partitions. GPU-serving pods consume fewer CPU cycles ({42--72m}) but require more memory ({773--876M}), while CPU pods demand more CPU ({715--1062m}) with slightly lower memory usage. All pods are scheduled within {1 second}, demonstrating efficient resource allocation under moderate load. However, the lack of dynamic scaling or predictive behavior underlines the need for adaptive autoscaling policies.

\begin{table}[htbp]
\centering
\caption{Baseline Resource Utilization and Scheduling (Sched.) Latency}
\label{tab:resource-utilization}
\begin{tabularx}{\linewidth}{X|c|c|c|c}
\toprule
\textbf{Pod Type} & \textbf{CPU} & \textbf{Memory} & \textbf{Replicas} & \textbf{Sched. Latency} \\
\midrule
GPU    & \textbf{42--72m}     & \textbf{773--876M}  & 3 & $<$1s \\
CPU    & \textbf{715--1062m}  & 534--550M  & 3 & $<$1s \\
Memory & 2--3m       & 3Mi         & 3 & $<$1s \\
\bottomrule
\end{tabularx}
\end{table}

\subsection{KIScaler Training and Learning Analysis}

We train {KIScaler}, a Proximal Policy Optimization (PPO)-based reinforcement learning agent, for 100 episodes, where each episode corresponds to a full simulated inference session using a rotating traffic pattern. The training environment is provided by {KISim}, which delivers fine-grained, GPU-aware feedback at each decision step. The reward function combines P95 latency minimization, GPU utilization maximization, and scaling penalty reduction, guiding the agent to learn balanced, cost-efficient autoscaling strategies.

\begin{figure}[htbp]
\centering
\begin{subfigure}[t]{0.8\linewidth}
    \centering
    \includegraphics[width=\linewidth]{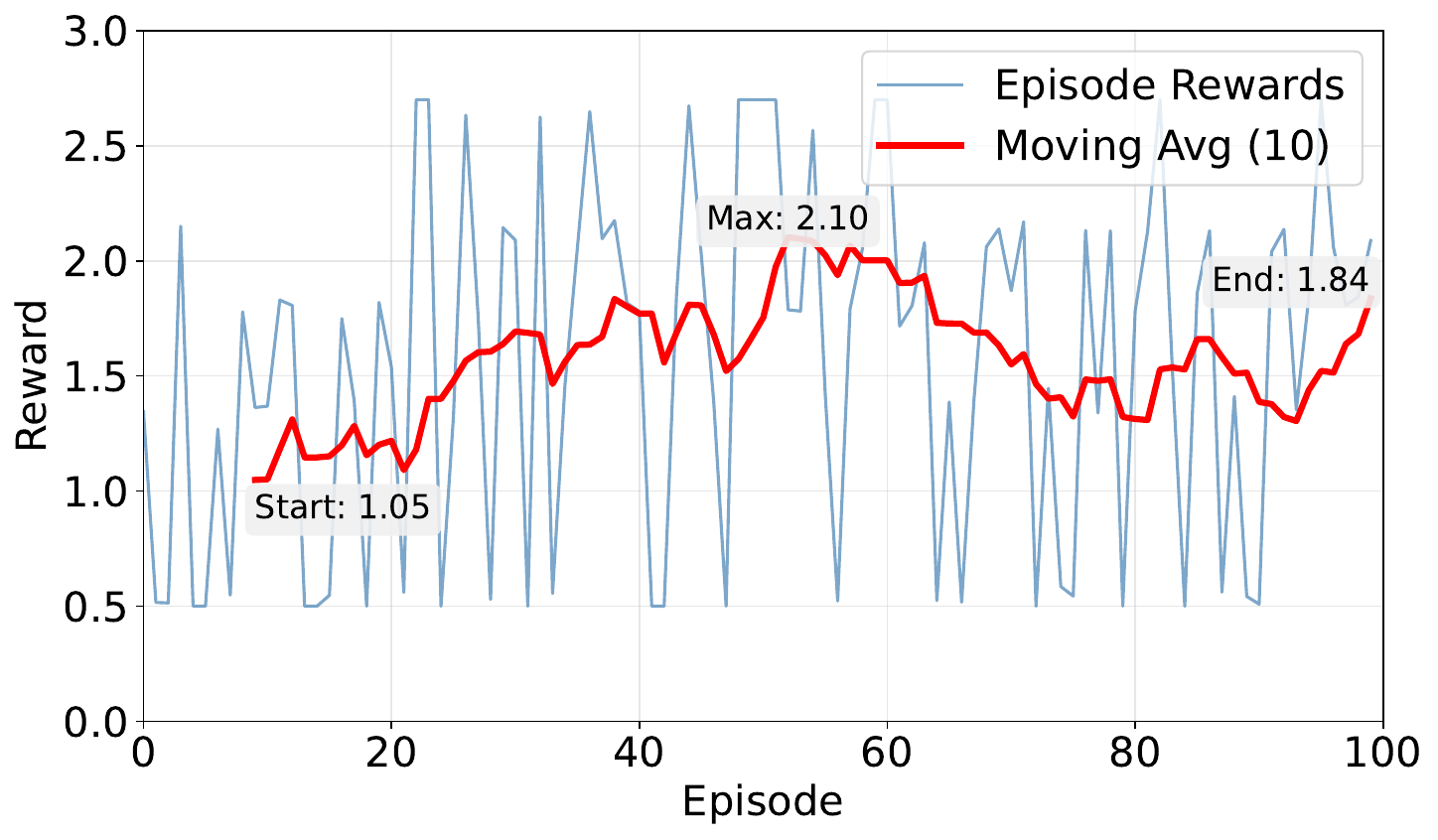}
    \caption{Training rewards and moving average.}
\end{subfigure}
\vskip 0.5em
\begin{subfigure}[t]{0.8\linewidth}
    \centering
    \includegraphics[width=\linewidth]{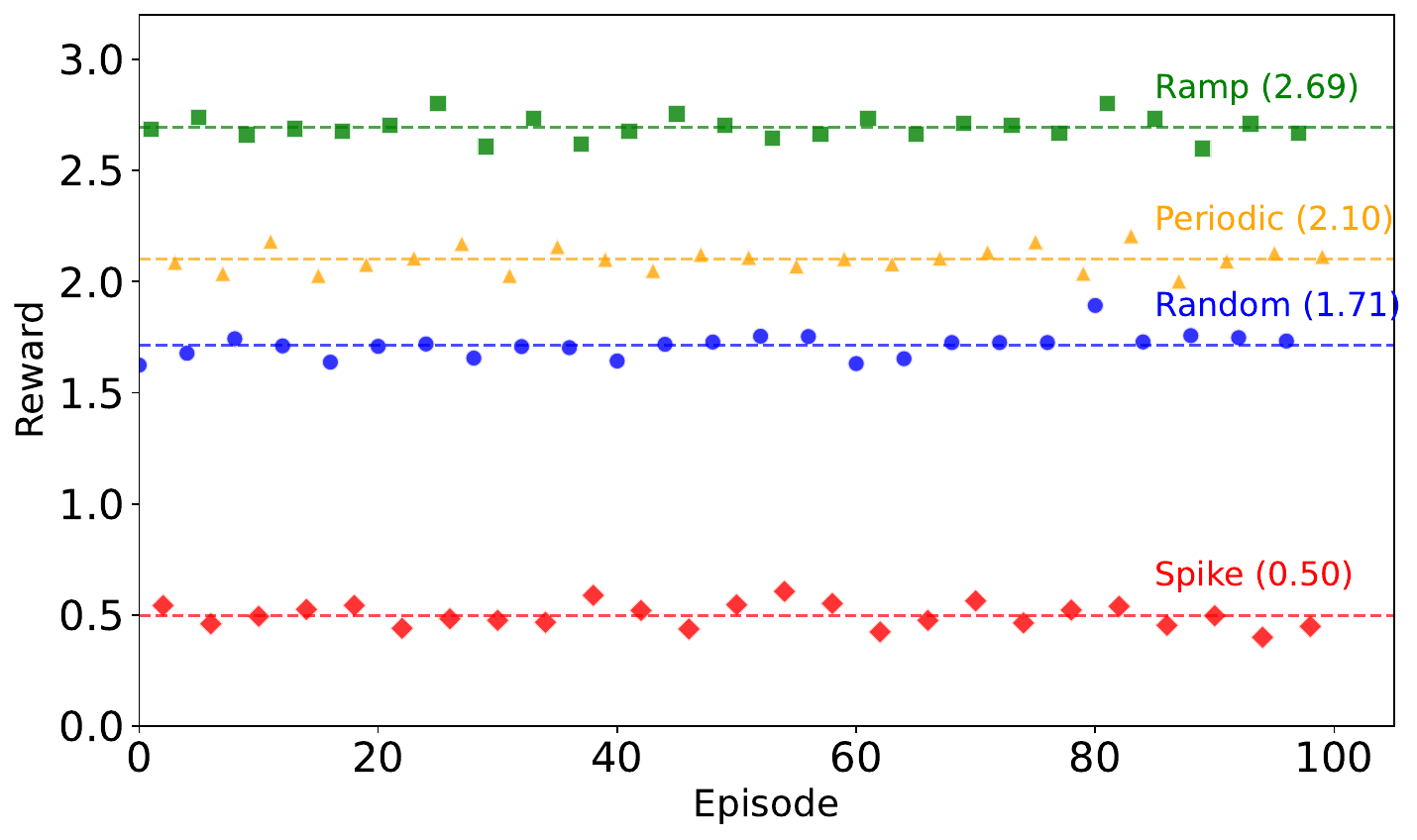}
    \caption{Pattern-specific reward trends across episodes.}
\end{subfigure}
\vskip 0.5em
\begin{subfigure}[t]{0.8\linewidth}
    \centering
    \includegraphics[width=\linewidth]{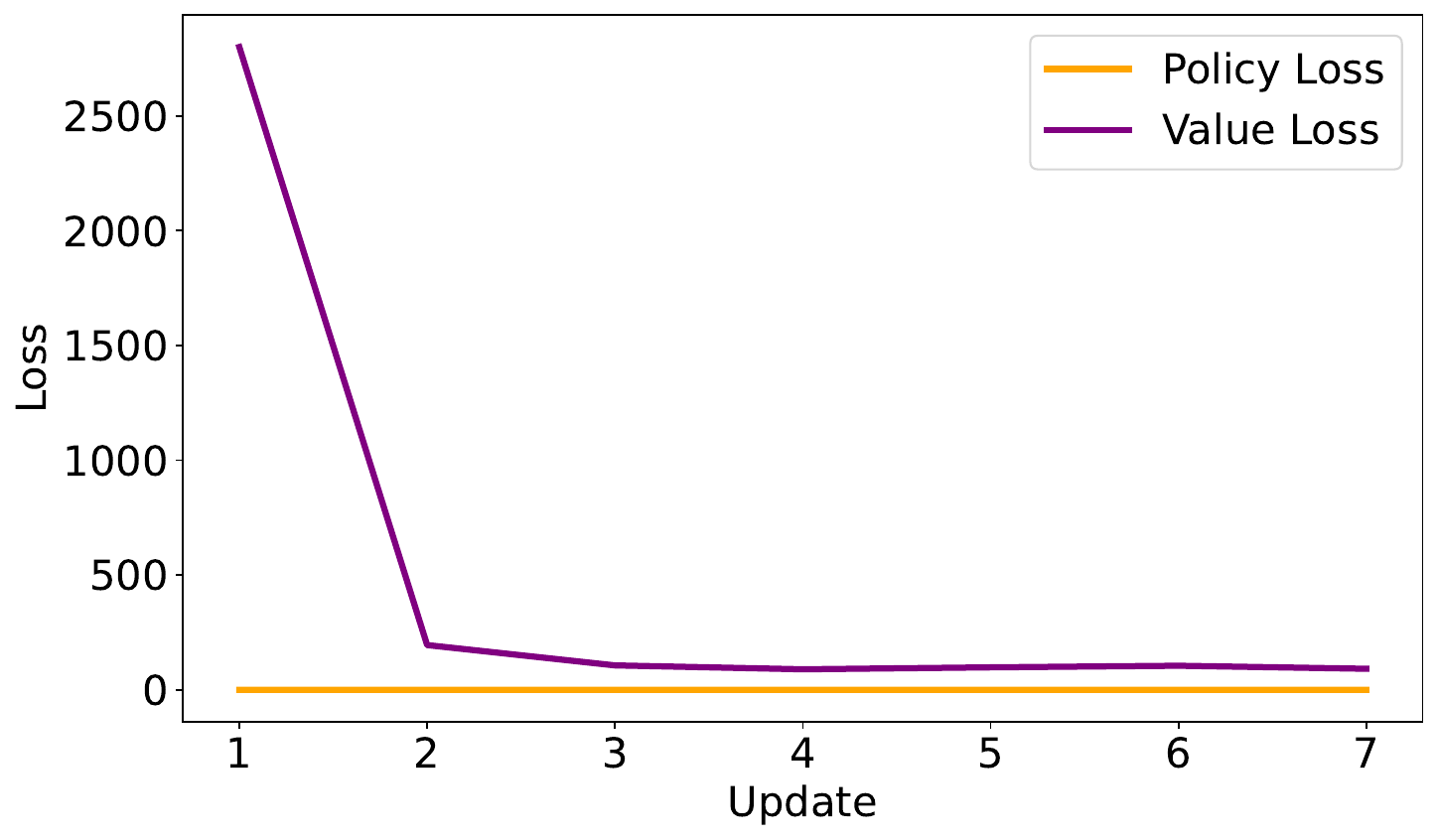}
    \caption{Convergence of policy and value losses.}
\end{subfigure}
\caption{Reinforcement learning progress of KIScaler.}
\label{fig:training_progress}
\end{figure}

Fig.~\ref{fig:training_progress} shows the RL progress of KIScaler. From Fig.~\ref{fig:training_progress}, we observe that KIScaler adapts to diverse patterns and consistently improves reward, even in highly stochastic environments like \textit{random} and \textit{spike}.

Fig.~\ref{fig:training_progress}(a) shows the episodic reward progression throughout training. The agent starts with an initial moving average reward of {1.05}, gradually improves over time, and reaches {1.84} by the final training episodes—a {75.2\% improvement}. The curve peaks at {2.10}, reflecting strong policy performance under favorable traffic conditions. This consistent upward trend confirms stable and effective learning of the autoscaling policy through interaction with the simulated environment.

Fig.~\ref{fig:training_progress}(b) breaks down the learning performance by traffic pattern. We observe that \textit{periodic} and \textit{ramp} patterns yield the highest and most consistent rewards due to their predictability, enabling the agent to anticipate scaling needs. The \textit{random} pattern falls in between, showing decent performance with occasional dips. Specially, the \textit{spike} pattern presents the most learning difficulty, resulting in lower mean reward, likely due to delayed scaling responses during sudden bursts.

Fig.~\ref{fig:training_progress}(c) shows the training loss curves for the PPO agent. The value loss (purple) starts high—typical of a randomly initialized critic—but quickly drops, indicating rapid improvement in value estimation as the agent gains experience. The policy loss (orange) remains low and stable, reflecting PPO’s ability to constrain updates within a safe region and prevent instability. Together, the two curves suggest efficient and stable learning: the critic converges quickly, and the actor refines its scaling strategy progressively without erratic shifts.

Taken together, these trends demonstrate that KIScaler successfully learns to distinguish between different load patterns and tailors its autoscaling behavior accordingly. The agent converges to a stable, high-performing policy that generalizes across dynamic inference workloads—an essential property for real-world deployment in unpredictable cloud environments.

\begin{comment}

\begin{figure*}[htbp]
\centering
\begin{subfigure}[t]{0.44\linewidth}
    \centering
    \includegraphics[width=\linewidth]{figs/rl/rl_training_progress.pdf}
    \caption{RL Training Progress}
\end{subfigure}
\begin{subfigure}[t]{0.44\linewidth}
    \centering
    \includegraphics[width=\linewidth]{figs/rl/training_losses.pdf}
    \caption{Training Losses}
\end{subfigure}
\caption{RL Training Progress: Episode rewards with moving average trend.}
\label{fig:rl-training-progress}
\end{figure*}

\begin{figure*}[htbp]
\centering
\begin{subfigure}[t]{0.44\linewidth}
    \centering
    \includegraphics[width=\linewidth]{figs/rl/rl_learning_by_pattern.pdf}
    \caption{Learning Progress by Pattern}
\end{subfigure}
\begin{subfigure}[t]{0.44\linewidth}
    \centering
    \includegraphics[width=\linewidth]{figs/rl/p95_latency_comparison.pdf}
    \caption{}
\end{subfigure}
\caption{}
\label{fig:rl-analysis}
\end{figure*}
\end{comment}

\subsection{KIScaler Outperforms Static Baselines}

\begin{figure}[htbp]
    \centering
    \includegraphics[width=0.85\linewidth]{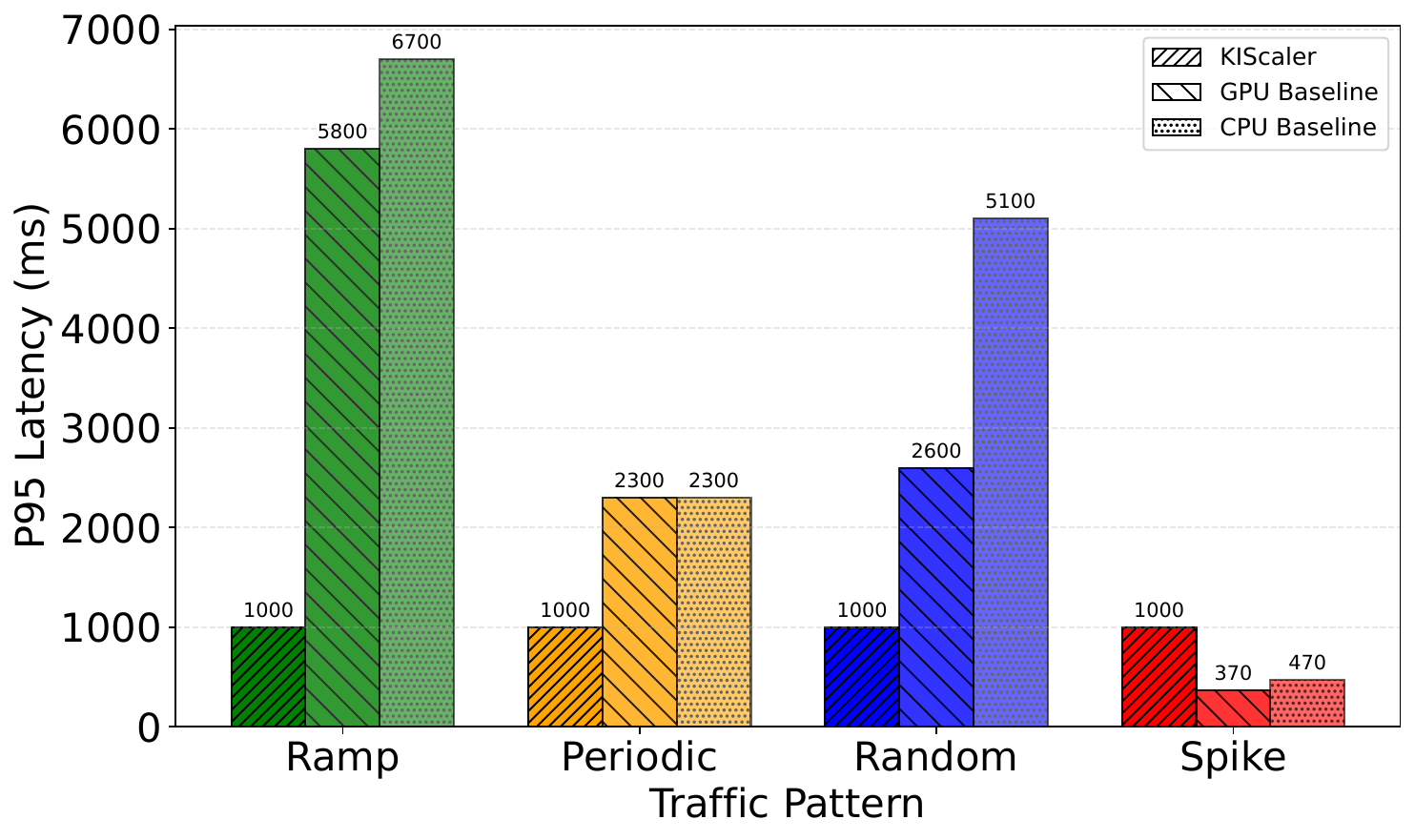}
    \caption{P95 latency comparison across traffic patterns for KIScaler, GPU baseline, and CPU baseline.}
    \label{fig:p95-latency-comparison}
\end{figure}

Table~\ref{tab:rl-baseline-comparison} and Fig.~\ref{fig:p95-latency-comparison} compare the P95 latency of {KIScaler} against static GPU and CPU baselines under four synthetic traffic patterns. Despite being trained solely in simulation, KIScaler consistently outperforms both baselines in dynamic settings.

Under \textit{ramp} traffic, KIScaler achieves the largest improvement, reducing latency by up to {6.7$\times$} compared to the CPU baseline. For the \textit{random} pattern, characterized by unpredictability, it still achieves {2.6$\times$ / 5.1$\times$} speedups over GPU and CPU deployments, respectively. In the \textit{periodic} scenario, the gains are consistent ({2.3$\times$}) across both baselines, showing KIScaler's ability to exploit predictable cycles. While in the \textit{spike} case, baseline latencies are already low, leaving less room for improvement.

These results demonstrate KIScaler's generalization ability across diverse traffic dynamics and highlight its advantage over threshold-based autoscaling under nonstationary workloads.

\begin{table}[htbp]
\centering
\caption{P95 Latency Comparison: KIScaler vs. Baselines}
\label{tab:rl-baseline-comparison}
\resizebox{\linewidth}{!}{%
\begin{tabular}{l|c|c|c|c}
\toprule
\textbf{Pattern} & \textbf{KIScaler (ms)} & \textbf{GPU Baseline} & \textbf{CPU Baseline} & \textbf{Speedup} \\
\midrule
Ramp     & \textbf{1000} & 5800 & 6700 & 5.8$\times$ / 6.7$\times$ \\
Periodic & \textbf{1000} & 2300 & 2300 & 2.3$\times$ / 2.3$\times$ \\
Random   & \textbf{1000} & 2600 & 5100 & 2.6$\times$ / 5.1$\times$ \\
Spike    & \textbf{1000} & 370  & 470  & 0.37$\times$ / 0.47$\times$ \\
\bottomrule
\end{tabular}}
\end{table}

%%%%%%%%%%%%%%%%%%%%%%%%%%%%%%%%%%%%%%%%%%%%%%%%%%%%%%%%%%%%%%%%%%%%%%%%%%%%%%%%%%%%%%%%% Section Discussion and Future work                      %%%%%%%%%%%%%%
%%%%%%%%%%%%%%%%%%%%%%%%%%%%%%%%%%%%%%%%%%%%%%%%%%%%%%%%%%%%%%%%%%%%%%%%%%%%%%%%
\section{Discussion}
\label{sec:discussion}

Our experiments highlight the feasibility and benefits of RL-based autoscaling in dynamic, GPU-enabled Kubernetes environments. Trained entirely in simulation, KIScaler learns pattern-specific scaling strategies and consistently outperforms threshold-based baselines such as HPA.

\mypara{Interpretation of Key Findings} KIScaler exhibits strong learning signals across all evaluated traffic patterns. It performs best under predictable loads like \textit{periodic} and \textit{ramp}, where it learns to anticipate scaling demands. Under more volatile workloads such as \textit{spike} and \textit{random}, the agent shows moderate performance with increased reward variance. The presence of multimodal reward distributions suggests that KIScaler develops specialized policies for different load types, rather than converging on a single generic scaling heuristic.

\mypara{System Constraints and Deployment Limitations} Our prototype was deployed on a single-node cluster with one NVIDIA RTX 3080 GPU, limiting concurrent GPU pod execution. While all CPU pods could run in parallel, only one GPU-based pod was schedulable at a time. This introduced resource imbalance and restricted the RL agent's action space. To address this, we used synthetic feedback from the simulator during training, which enabled safe, accelerated learning and fallback exploration. Despite these hardware constraints, KIScaler successfully learned effective scaling policies and generalized to real cluster settings without retraining.

\mypara{Future Directions} We plan to scale KIScaler to multi-node clusters with multiple GPUs to better represent production-scale deployments. Future iterations will replace synthetic simulation signals with live inference metrics, enabling online learning and adaptation. Additionally, we aim to enhance the reward function to incorporate QoS-aware and fairness-driven objectives, aligning with the needs of latency-sensitive, multi-tenant AI workloads.

%%%%%%%%%%%%%%%%%%%%%%%%%%%%%%%%%%%%%%%%%%%%%%%%%%%%%%%%%%%%%%%%%%%%%%%%%%%%%%%%%%%%%%%%%
\section{Related Work}
\label{sec:related_work}

\subsection{Autoscaling in GPU-Accelerated Kubernetes Environments}

Kubernetes' default HPA scales workloads based on CPU and memory thresholds but is poorly suited for GPU-accelerated inference, which often involves bursty, latency-sensitive workloads~\cite{han2022microsecond, ahmad2024smart}. To address these limitations, Smart HPA~\cite{ahmad2024smart} introduces a hierarchical, resource-efficient control plane, while ProSmart HPA~\cite{singh2022prosmart} adds a proactive, machine learning-based policy to mitigate pod initialization delays. Trend-aware autoscalers~\cite{ahmad2025trendaware} further improve responsiveness by detecting short-term workload trends using predictive analytics.

SLO-driven frameworks such as LSRAM~\cite{hu2025lsram} formulate autoscaling as an optimization problem, enabling rapid resource allocation under changing loads. However, these approaches rely on fixed heuristics or offline-trained models, limiting adaptability to unseen traffic patterns or evolving QoS constraints. In contrast, reinforcement learning (RL) offers a principled way to learn adaptive, multi-objective scaling policies through real-time system interaction~\cite{mao2016resource_management_rl}.

Despite RL’s potential, its adoption remains limited due to the lack of reproducible, GPU-aware simulation environments. While KubeShare~\cite{yeh2020kubeshare} enhances GPU sharing in Kubernetes by treating GPUs as first-class schedulable resources, it does not support autoscaling or policy learning. Our framework addresses this gap by combining high-fidelity GPU simulation (KISim) with an RL-based autoscaler (KIScaler) that is trainable and deployable in real Kubernetes clusters.
%%%%%%%%%%%%%%%%%%%%%%%%%%%%%%%%%%%%%%%%%%%%%%%%%%%%%%%%%%%%%%%%%%%%%%%%%%%%%%%%%%%%%%%%% Section Conclusion                          %%%%%%%%%%%%%%
%%%%%%%%%%%%%%%%%%%%%%%%%%%%%%%%%%%%%%%%%%%%%%%%%%%%%%%%%%%%%%%%%%%%%%%%%%%%%%%%
\section{Conclusion}
\label{sec:conclusion}

We present \textbf{KIS-S}, a unified framework for intelligent autoscaling of GPU-accelerated inference workloads in Kubernetes, addressing the limitations of traditional threshold-based strategies. KIS-S consists of two key components: \textbf{KISim}, a GPU-aware simulator for safe and reproducible training, and \textbf{KIScaler}, a reinforcement learning (RL)-based autoscaler trained via Proximal Policy Optimization (PPO) to learn adaptive scaling policies.

We evaluate the system using four synthetic load patterns—ramp, spike, periodic, and random—serving real MobileNetV4 inference through NVIDIA Triton. Baseline experiments show that GPU inference outperforms CPU in bursty or irregular workloads, achieving up to 1.96$\times$ speedup. However, the default Kubernetes scheduler fails to adapt to workload variability.

KIScaler, trained under resource-constrained conditions with synthetic feedback, improves its final moving average reward by {75.2\%} over the course of training and exhibits robust, pattern-specific behavior. It reduces P95 latency by up to {6.7$\times$} compared to CPU baselines. KISim enables safe and flexible experimentation, while KIScaler integrates directly with Kubernetes APIs and Prometheus for closed-loop control.

This work demonstrates the feasibility and efficacy of deploying RL-based autoscalers in GPU-accelerated Kubernetes environments, offering not only a practical path toward intelligent orchestration beyond traditional threshold-based methods, but also laying a foundation for scalable, adaptive resource management in real-world AI inference systems where workload dynamics and latency constraints are critical.

\bibliographystyle{IEEEtran}
\bibliography{reference}

\end{document}